\documentclass{aa}
\usepackage{graphics,epsfig,txfonts}

\usepackage{natbib}
\bibpunct{(}{)}{;}{a}{}{,}

\newcommand{\ergcm}[1]{$\times 10^{#1}$ erg cm$^{-2}$ s$^{-1}$}
\newcommand{\oergcm}[1]{$10^{#1}$ erg cm$^{-2}$ s$^{-1}$}
\newcommand{\ergs}[1]{$\times 10^{#1}$ erg s$^{-1}$}

\newcommand{\hcm}[1]{$\times 10^{#1}$ cm$^{-2}$}
\newcommand{\ohcm}[1]{$10^{#1}$ cm$^{-2}$}
\newcommand{\expo}[1]{$\times 10^{#1}$}

\newcommand{\kms}{km s$^{-1}$}
\newcommand{\nh}{N$_{\rm H}$}

\newcommand{\lx}{\hbox{L$_{\rm x}$}}
\newcommand{\lopt}{\hbox{L$_{\rm opt}$}}

\newcommand{\Hone}{\ion{H}{I}}

\newcommand{\ltsima}{$\buildrel < \over \sim$}
\newcommand{\lsim}{\lower.5ex\hbox{\ltsima}}
\newcommand{\gtsima}{$\buildrel > \over \sim$}
\newcommand{\gsim}{\lower.5ex\hbox{\gtsima}}
\newcommand{\sn}{\hbox{SN\,1987\,A}}

\begin{document}
 
\title{XMM-Newton observations of \object{SN\,1987\,A}
}
 
\author{F.~Haberl \and U.~Geppert \and B.~Aschenbach \and G.~Hasinger}

\titlerunning{XMM-Newton observations of SN\,1987\,A}
\authorrunning{Haberl et al.}
 
\offprints{F. Haberl, \email{fwh@mpe.mpg.de}}
 
\institute{Max-Planck-Institut f\"ur extraterrestrische Physik,
           Giessenbachstra{\ss}e, 85748 Garching, Germany
	   }
 
\date{Received 7 August 2006 / Accepted 13 September 2006}
 
\abstract{We report on XMM-Newton observations of \sn\ in the Large Magellanic Cloud.}
         {The large collecting area telescopes together with the European Photon Imaging 
	  Cameras (EPIC) provide X-ray spectra with unprecedented statistical quality and 
	  make it possible to investigate the spectral evolution during the brightening 
	  observed since the discovery in X-rays. High resolution spectra from the 
	  Reflection Grating Spectrometers yield a complementary view and allow us to 
	  perform more detailed investigations of prominent emission lines.}
	 {The X-ray spectra were modeled with two-temperature emission components
	  from a hot plasma in collisional ionization equilibrium and in non-equilibrium 
	  (NEI).}
         {We find a temperature for the equilibrium component of 0.24$\pm$0.02 keV in
	  January 2000 and April 2001 which increased to 0.30$\pm$0.02 keV in May 2003 and also
	  an indication for a temperature increase in the hot NEI component from $\sim$2 keV
	  to $\sim$3 keV. Emission line ratios inferred from the RGS spectra suggest
	  temperatures as low as 100 eV and an increase in the ionization state of
	  oxygen and neon consistent with the observed temperature increases.
	  The fast readout of the EPIC-pn instrument yields X-ray fluxes free of CCD 
	  pile-up effects which we used to normalize pile-up corrections for the published
	  Chandra fluxes. The corrected X-ray light curve of \sn\ in the 0.5-2.0 keV energy 
	  band is best represented by a linear increase up to about day 4000 after the 
	  explosion and an exponential rise afterwards until the last published Chandra
	  observation on day 6716. Modeling the light curve by emission from the inner ring 
	  which is approximated by a circular torus a central density 
	  n$\sb{\rm{H}}$~=~1.15$\times$10$\sp 4$~cm$\sp{-3}$ is found. In this model the 
	  forward shock has just passed the center of the torus.}
         {\sn\ continues to brighten exponentially in soft X-rays. The X-ray spectra can be 
	  represented by pure thermal emission without significant contribution from a compact
	  object yet.}

 
\maketitle
 
\section{Introduction}

The first supernova observed in 1987 (\sn) went off in the Large Magellanic Cloud (LMC) on 
23 February. This event with observations involving the entire electromagnetic spectrum and the first 
detection of cosmic neutrinos brought about a number of astrophysical surprises, some of which 
are puzzling still today. In this paper we concentrate on the observations of soft (0.1 - 3 keV) 
X-ray emission since the explosion, thought to arise from the interaction of shocks with the ambient 
circumstellar medium, and possibly a central compact object like a neutron star. 
We present and summarize for the first time the full set of previous data, which to some 
extent have been re-calibrated, and we add the XMM-Newton observations. 

The first attempts to search for soft X-rays were made in August 1987 with sounding rockets carrying 
imaging telescopes. An upper limit of 1.5$\times$10$\sp{36}$ erg~s$\sp{-1}$ could be established which 
excluded an ambient dense and slow wind typical for a red supergiant wind, but which is  
still compatible with a blue supergiant \citep{1987Natur.330..232A}.  
X-rays from \sn\ became clearly visible in the ROSAT images not much before spring 1992 
\citep{1994A&A...281L..45B,1994ApJ...420L..25G}. The monitoring of \sn\ with ROSAT revealed a 
slow, but steady increase in flux over the first years 
\citep{1996A&A...312L...9H}. The 
brightening continues since then but with increasing pace.  
\citep{2000ApJ...543L.149B,2002ApJ...567..314P,2004ApJ...610..275P,2005ApJ...634L..73P}. 
Chandra has taken images of \sn\ since late 1999, which show a ring-like emission region steadily 
growing in diameter. Details of the ring structure became apparent in the meantime. These structural features 
are correlated with the optical hot spots, first discovered in earlier 
images taken with the Hubble Space Telescope (HST). These optically bright spots evolve quite rapidly and they now seem to 
completely engulf the so-called inner ring  
\citep{2000ApJ...537L.123L,2005coex.conf...77M}. 

The ring with its fairly high matter density existed before the explosion and was 
detected by re-emitted light of the prompt supernova EUV flash. 
A substantial heating, up to X-ray temperatures, of the ring was predicted for 
the period between 2002 and 2010 when the supernova shock wave would have reached 
and entered the ring. Both the optical and X-ray images and the X-ray light curve 
demonstrate that this state has been arrived at. Hope is that we can learn more about 
the details of the interaction process, the origin of the ring as well as of the 
progenitor star and its evolution by X-ray spectroscopy. 

\begin{table*}
\caption[]{XMM-Newton observations of \sn\ and data selection.}
\begin{center}
\begin{tabular}{llcclrrr}
\hline\hline\noalign{\smallskip}
\multicolumn{1}{l}{Instrument} &
\multicolumn{1}{c}{Read-out} &
\multicolumn{1}{c}{Filter} &
\multicolumn{1}{c}{Sat. Revol.} &
\multicolumn{1}{c}{Date} &
\multicolumn{1}{c}{Time (UT)} &
\multicolumn{1}{c}{Net Exp.$^1$} &
\multicolumn{1}{c}{Counts$^2$} \\
\multicolumn{1}{c}{} &
\multicolumn{1}{c}{Mode} &
\multicolumn{1}{c}{} &
\multicolumn{1}{c}{} &
\multicolumn{1}{c}{} &
\multicolumn{1}{c}{} &
\multicolumn{1}{c}{[s]} &
\multicolumn{1}{c}{} \\

\noalign{\smallskip}\hline\noalign{\smallskip}
EPIC-pn      & FF, 73 ms & Medium &  21 & 2000 Jan. 19/20 & 16:19 -- 04:28 &  23250 &  2194 \\
EPIC-pn      & FF, 73 ms & Medium &  22 & 2000 Jan. 21    & 15:38 -- 19:34 &   5789 &   518 \\
EPIC-pn      & FF, 73 ms & Thin   &  22 & 2000 Jan. 21/22 & 20:32 -- 12:02 &  30990 &  3060 \\
\noalign{\smallskip}\hline\noalign{\smallskip}
EPIC-pn      & FF, 73 ms & Medium & 244 & 2001 Apr. 8/9   & 20:35 -- 07:14 &  24534 &  3952 \\
\noalign{\smallskip}\hline\noalign{\smallskip}
EPIC-pn      & SW, 6 ms  & Medium & 626 & 2003 May 10/11  & 16:41 -- 23:37 &  55855 & 25609 \\
EPIC-MOS1    & FF, 2.6 s & Medium &     &                 & 16:35 -- 23:37 & 101520 & 16020 \\
EPIC-MOS2    & FF, 2.6 s & Medium &     &                 & 16:35 -- 23:37 & 101510 & 16018 \\
RGS1         & Spectro   & --     &     &                 & 11:42 -- 23:38 & 112160 &  2509 \\
RGS2         & Spectro   & --     &     &                 & 10:47 -- 23:38 & 111670 &  3358 \\
\noalign{\smallskip}\hline
\end{tabular}
\end{center}

$^1$Net exposures for the observations from revolutions 21/22 were derived from a comparison
of the spectrum of N\,157\,B with that obtained from the ROSAT PSPC due to missing timing
information in the early data \citep{2001A&A...365L.208H}.\\
$^2$Net counts as used for the spectral analysis in the following energy bands: 
0.2$-$9.0 keV for the EPIC instruments 
(except for EPIC-pn during revolutions 21/22 when we adopted 0.3$-$9.0 keV due to the 
higher low-energy threshold used at the beginning of the performance verification phase); 
0.3$-$2.0 keV for RGS.
\label{xray-obs}
\end{table*}

The medium energy resolution X-ray spectrum measured by the Chandra ACIS detector is composed of two emission components with
temperatures kT of $\sim$2.4 keV and $\sim$0.22 keV, which sometimes are interpreted as contributions from the 
blast wave shock and the decelerated slow shock, respectively \citep{2002ApJ...574..166M,2004ApJ...610..275P}. The strong 
brightness increase made it also possible to obtain high resolution X-ray spectra with the Chandra LETG.
A total exposure of 289 ks was accumulated between August 26 and September 5, 2004.
Inferred temperatures from the emission line ratios range from 0.1 to 2 keV and expansion velocities are
found between 300 and 1700 \kms\ \citep{2005ApJ...628L.127Z}.

One other outstanding issue is the existence and the time of appearance of a central compact remnant like 
a neutron star. So far the results have been negative. The most stringent upper limits for an X-ray point source in the 
newly born supernova remnant are based on the analysis
of Chandra images in the 2--10 keV band with \lx\ $<$ 1.5\ergs{34} \citep{2004ApJ...610..275P} and 
on INTEGRAL measurements in the 20--60 keV band \citep[\lx\ $<$ 1.1\ergs{36}, ][]{2005AstL...31..258S}.
The HST derived optical limit is \lopt\ $<$ 5\ergs{33} \citep{2005ApJ...629..944G}.

XMM-Newton observed \sn\ on several occasions. Here, we report the results of the spectral analysis 
of the data with the highest statistical quality available to date, making use of the EPIC cameras and the RGS 
spectrometers.  

\section{X-ray observations}

XMM-Newton \citep{2001A&A...365L...1J} observations from three epochs were used to 
investigate the spectral evolution of \sn\ (for observation details see Table~\ref{xray-obs}).
The 30 Doradus region in the Large Magellanic Cloud was selected for first-light 
including \sn\ at an off-axis angle of about 7 arc minutes \citep{2001A&A...365L.202D}. 
Five exposures were taken with the EPIC-pn instrument \citep{2001A&A...365L..18S} 
in full-frame (FF) imaging mode which we use for spectral analysis as described by 
\citet{2001A&A...365L.208H}. Exposures with the same 
instrumental setup (pointing direction, optical blocking filter) were merged 
resulting in three EPIC-pn ``first-light" spectra of \sn\ from January 2000.
An observation at a second epoch in April 2001 was performed utilizing all instruments
on board of XMM-Newton; the CCD cameras EPIC-pn and EPIC-MOS \citep{2001A&A...365L..27T}
and the Reflection Grating Spectrometers \citep[RGS, ][]{2001A&A...365L...7D}.
For spectral analysis we use the EPIC-pn data from this observation. 

Spectra from all instruments were accumulated from the $\sim$100 ks observation
performed in May 2003 when the source was sufficiently bright such that the long exposure yielded 
spectra with sufficient statistical quality for all instruments. 
During this observation EPIC-pn was set to the faster small-window (SW) CCD readout mode.
For optical light blocking the ``medium'' filter was selected in the EPIC cameras, 
except during the last part of the first-light observations, when a thin 
filter was used. 

We screened the data for high background intervals and extracted source and 
background spectra of \sn\ from circular regions with a  
25\arcsec\ radius placed on the source and a nearby point-source free area.
For the EPIC-pn spectra single-pixel (PATTERN 0) events were selected while for EPIC-MOS
all valid event patterns (PATTERN 0 to 12) were used.
The spectra were binned to contain a minimum of 30 and 40 counts per bin for EPIC and RGS,
respectively. Table~\ref{xray-obs} lists net exposures and net counts accumulated for 
the spectra.
The data from satellite revolutions 244 and 626 were processed with SAS release 6.1.0
together with the latest calibration files relevant for spectral analysis of EPIC-pn data
(EPN\_REDIST\_0010.CCF for the re-distribution matrices and XRT3\_XAREA\_0010.CCF with an 
update to the mirror effective area).
The spectra from revolutions 21/22 were directly taken from the work of 
\citet{2001A&A...365L.202D} and \citet{2001A&A...365L.208H} together with new response files mentioned 
above.
Spectral fitting was performed using XSPEC \citep{1996ASPC..101...17A} version 11.3.2i.

\begin{table*}
\caption[]{Fit quality of two-temperature models.}
\begin{center}
\begin{tabular}{lrrrrrrrrr}
\hline\hline\noalign{\smallskip}
\multicolumn{1}{l}{Model$^1$} &
\multicolumn{3}{c}{EPIC-pn, all} &
\multicolumn{3}{c}{MOS, rev. 626} &
\multicolumn{3}{c}{RGS, rev. 626} \\
\multicolumn{1}{l}{} &
\multicolumn{1}{c}{$\chi^2$} &
\multicolumn{1}{c}{dof} &
\multicolumn{1}{c}{$\chi^2_r$} &
\multicolumn{1}{c}{$\chi^2$} &
\multicolumn{1}{c}{dof} &
\multicolumn{1}{c}{$\chi^2_r$} &
\multicolumn{1}{c}{$\chi^2$} &
\multicolumn{1}{c}{dof} &
\multicolumn{1}{c}{$\chi^2_r$}\\

\noalign{\smallskip}\hline\noalign{\smallskip}
A: PHABS * VPHABS * (VNEI + VRAYMOND) &  806.5 & 716 & 1.126 & 864.1 & 357 & 2.434 & 799.0 & 580 & 1.377 \\
B: PHABS * VPHABS * (VNEI + VMEKAL)   &  804.8 & 716 & 1.124 & 903.6 & 357 & 2.545 & 811.4 & 580 & 1.399 \\
C: PHABS * VPHABS * (VNEI + VNEI)     &  802.2 & 715 & 1.122 & 890.4 & 357 & 2.508 & 870.2 & 580 & 1.500 \\
D: PHABS * VPHABS * (VMEKAL + VMEKAL) &  858.1 & 717 & 1.197 & 878.4 & 355 & 2.474 & 810.3 & 580 & 1.397 \\
\noalign{\smallskip}\hline
\end{tabular}
\end{center}

$^1$Model components as available in XSPEC: PHABS represents the Galactic foreground absorption, VPHABS the absorption
in the LMC with reduced metal abundances, VRAYMOND and VMEKAL emission from hot plasma in collisional ionization 
equilibrium with variable abundances and VNEI emission from a non-equilibrium plasma (see also Sect. 3).
\label{chisq}
\end{table*}

\begin{figure*}
  \resizebox{12cm}{!}{\includegraphics[angle=-90,clip=]{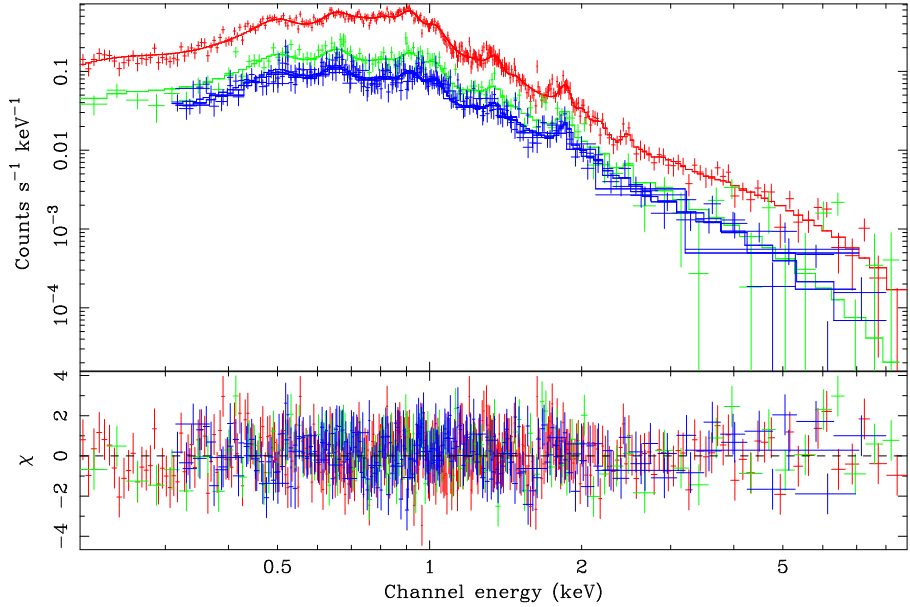}}
  \hfill\parbox[t]{60mm}{
  \vspace{60mm}\caption[]{
      EPIC-pn spectra of \sn\ from Jan. 2000, April 2001 and May 2003 
      (in increasing intensity). The histograms show the best fit model A 
      as defined in Table~\ref{chisq} and parameters listed in 
      Tables~\ref{pn-fits-a} and \ref{pn-fits-b}.
  }
  \label{epicpn-spectra}
  }
\end{figure*}

\section{Spectral analysis}

For studying the spectral evolution of \sn\ we concentrate on the EPIC-pn data because they provide 
the spectra with the  highest statistical quality, in particular for the early observations when the source
was still comparatively faint. 
We simultaneously fit the five EPIC-pn spectra with different two-temperature 
models similar to those applied to Chandra ACIS spectra by \citet{2004ApJ...610..275P}.
The best results were found using a two component thermal model (c.f. Table~\ref{chisq}) involving both the collisional ionization equilibrium models   
 \citep[XSPEC models MEKAL or RAYMOND, ][]{1985A&AS...62..197M,1977ApJS...35..419R}  
and/or the non-equilibrium model \citep[NEI, ][]{2001ApJ...548..820B}. For the photo-electric foreground absorption 
we use a fixed Galactic column density of \nh\ = 6\hcm{20} with the elemental abundances given by 
\citet{2000ApJ...542..914W} and an LMC column density (free in the fit) with the same abundances relative to H but 
reduced by a factor of 2 
for each element except He (also relative to the Wilms et al. abundances). 
The, in comparison with the solar value of \citet{1989GeCoA..53..197A},  
lower Oxygen abundance of Wilms et al. better fits the EPIC spectra around the Oxygen K absorption edge. 

We allowed some key parameters of the models to vary with time, i.e. they are free parameters for the
individual spectra of different epochs. These are temperature and normalization (given in the 
following as emission measure). The ionization timescale in the NEI models was found to be the same within the 
error limits for the various spectra and therefore it was used just as one single free parameter
in the combined fit. Similarly, we assumed that the set of elemental abundances used in the emission models  
did not change with time as a whole, but we still fit abundances individually for each element. Also for the emission models 
abundances relative to those given by \citet{2000ApJ...542..914W} are used. For the three first-light spectra only 
a constant factor was allowed to vary to account for the uncertainties in the net exposure time. However
the spectra agree within $\pm$3\%.

\begin{figure*}
  \resizebox{12cm}{!}{\includegraphics[angle=-90,clip=]{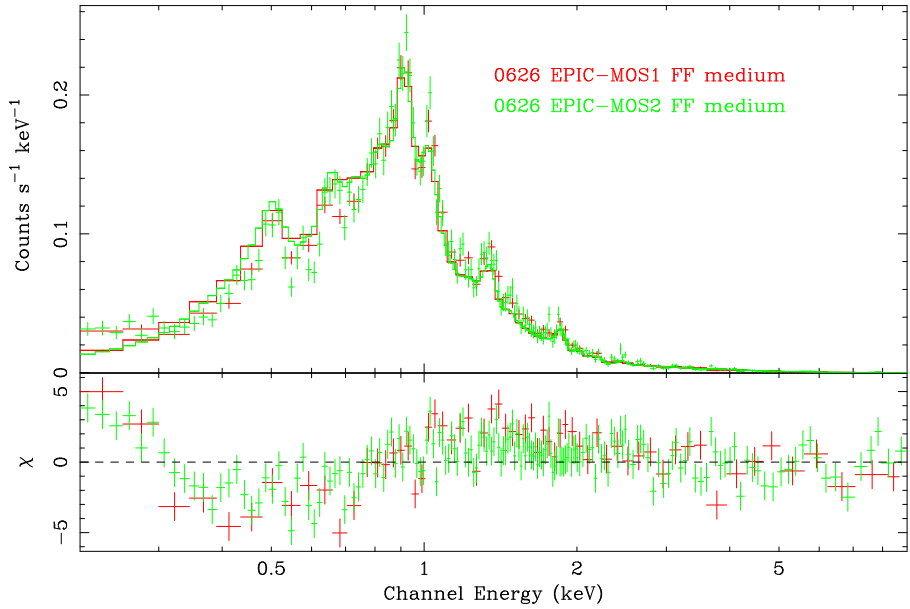}}
  \hfill\parbox[t]{60mm}{
  \vspace{54mm}\caption[]{
      EPIC-MOS spectra of \sn\ from satellite revolution 626. The histograms show 
      the prediction of model A with the best-fit parameter values of the EPIC-pn 
      spectrum except the normalization (a constant scaling factor of 0.98 and 
      0.99 for MOS1 and MOS2, respectively).
   }
   \label{epicmos-spectra}
   }
\end{figure*}

\begin{figure*}
  \resizebox{12cm}{!}{\includegraphics[angle=-90,clip=]{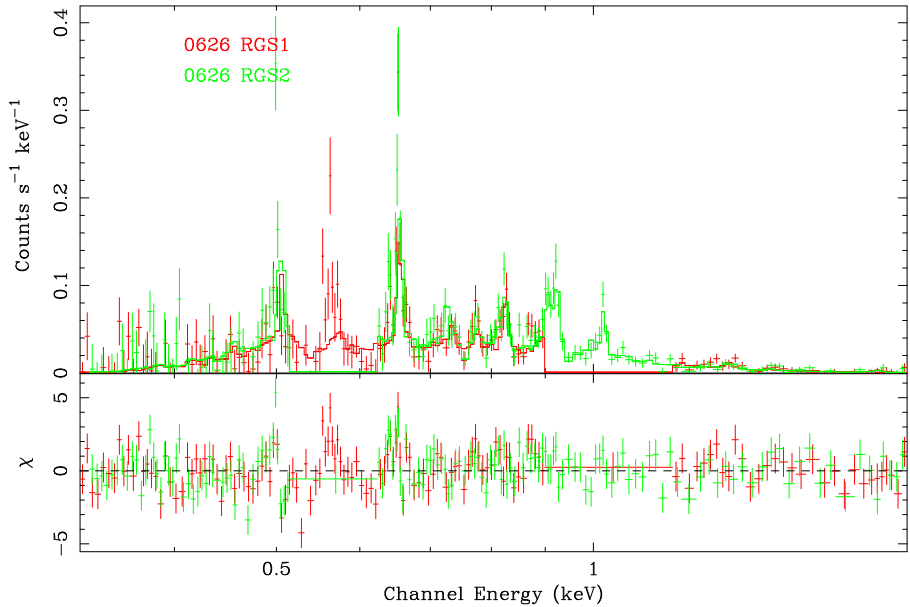}}
  \hfill\parbox[t]{60mm}{
  \vspace{64mm}\caption[]{
      RGS spectra of \sn\ from satellite revolution 626. The model is as 
      for Fig.\ref{epicmos-spectra} with scaling factors 1.01 and 0.98 for 
      RGS1 and RGS2, respectively.
  }
  \label{rgs-spectra}
  }
\end{figure*}

The energy resolution of the EPIC instruments does not allow to differentiate between the different
two-temperature models. Table~\ref{chisq} summarizes the quality of the fits for the best four models.
The combined EPIC-pn fits yield a somewhat smaller reduced $\chi^2$ values for models A, B and C which each involve a
non-equilibrium model for the high energy part of the spectrum. The EPIC-pn spectrum together with the best
fit model A is shown in Fig.~\ref{epicpn-spectra}. We cross-checked the results from
the EPIC-pn fits with the EPIC-MOS and RGS spectra from revolution 626 (earlier spectra have insufficient 
statistics) by folding the model through the corresponding detector response (SAS 6.1.0 version) but 
optimizing the normalization through a 
constant factor. The MOS spectra (Fig.~\ref{epicmos-spectra}) are fully compatible with
the pn results above $\sim$0.8 keV. Below that energy the MOS fits suffer from the changing event 
redistribution which is not included in the calibration files of SAS 6.1.0. The RGS spectra with their
higher energy resolution (Fig.~\ref{rgs-spectra}) formally exclude model C (this model produces a strong 
line at 0.512 keV which does not exist in the data).


In the following we present more detailed results for models A and B. The best fit parameters 
are summarized in Tables~\ref{pn-fits-a} and \ref{pn-fits-b}. Model A is a combination of NEI and RAYMOND emission 
components while model B uses NEI and MEKAL. The medium resolution EPIC-pn spectra can not
discriminate between these two models, but we prefer RAYMOND rather than MEKAL because the latter produces a
\ion{S}{XIV} line at 0.512 keV which is not seen in the data and not present in any other model (see RGS 
spectra in Fig.~\ref{rgs-spectra}). 
Apart from this difference the best fit parameters of model A 
and model B agree very well within their error limits.

\begin{table}[t]
\caption[]{Results from the two-temperature model fits to the EPIC-pn spectra
           - the parameters allowed to vary with time. Fluxes are given in
	   units of \oergcm{-13} and emission measures EM in 10$^{58}$ cm$^{-3}$.}
\begin{center}
\begin{tabular}{lccc}
\hline\hline\noalign{\smallskip}
\multicolumn{1}{l}{Parameter} &
\multicolumn{1}{c}{Rev. 626} &
\multicolumn{1}{c}{Rev. 244} &
\multicolumn{1}{c}{Rev. 21/22}\\
\multicolumn{1}{l}{} &
\multicolumn{1}{c}{May 2003} &
\multicolumn{1}{c}{Apr. 2001} &
\multicolumn{1}{c}{Jan. 2000}\\

\noalign{\smallskip}\hline\noalign{\smallskip}
 \multicolumn{4}{l}{Model A} \\ \vspace{1mm}
 NEI kT (keV)			  & 3.22$_{-0.49}^{+0.53}$ & 1.76$_{-0.34}^{+0.68}$  & 2.10$_{-0.29}^{+0.35}$ \\  \vspace{1mm}
 NEI EM                 	  & 1.00$_{-0.15}^{+0.08}$ & 0.69$_{-0.15}^{+0.18}$  & 0.43$_{-0.06}^{+0.06}$ \\  \vspace{1mm}
 RAYMOND kT (keV)		  & 0.30$_{-0.02}^{+0.02}$ & 0.24$_{-0.02}^{+0.02}$  & 0.24$_{-0.02}^{+0.02}$ \\  \vspace{1mm}
 RAYMOND EM                       & 16.0$_{-3.7}^{+3.0}$   & 5.1$_{-1.4}^{+1.1}$     & 3.2$_{-0.9}^{+0.7}$    \\
 Flux (0.5-2.0 keV)		  & 8.10		   & 2.77		     & 1.75		      \\
 Flux (3.0-10.0 keV)		  & 1.77		   & 0.43		     & 0.39		      \\
 Flux (0.2-10.0 keV)		  & 11.4		   & 3.82		     & 2.59		      \\
\noalign{\smallskip}\hline\noalign{\smallskip}
 \multicolumn{4}{l}{Model B} \\ \vspace{1mm}
 NEI kT (keV)			  & 3.11$_{-0.45}^{+0.63}$ & 1.65$_{-0.18}^{+0.42}$  & 2.01$_{-0.28}^{+0.17}$ \\  \vspace{1mm}
 NEI EM                   	  & 1.03$_{-0.14}^{+0.14}$ & 0.74$_{-0.15}^{+0.25}$  & 0.45$_{-0.05}^{+0.07}$ \\  \vspace{1mm}
 MEKAL kT (keV) 		  & 0.31$_{-0.02}^{+0.02}$ & 0.25$_{-0.02}^{+0.03}$  & 0.24$_{-0.02}^{+0.01}$ \\  \vspace{1mm}
 MEKAL EM                         & 13.4$_{-1.3}^{+1.4}$   & 4.0$_{-0.9}^{+0.9}$     & 2.6$_{-0.6}^{+0.6}$    \\
 Flux (0.5-2.0 keV)		  & 8.11		   & 2.78		     & 1.75		      \\
 Flux (3.0-10.0 keV)		  & 1.74		   & 0.40		     & 0.38		      \\
 Flux (0.2-10.0 keV)		  & 11.4		   & 3.79		     & 2.58		      \\
\noalign{\smallskip}\hline
\end{tabular}
\end{center}
\label{pn-fits-a}
\end{table}
\begin{table}[t]
\caption[]{Results from the two-temperature model fits to the EPIC-pn spectra
           - the parameters assumed to be constant with time.}
\begin{center}
\begin{tabular}{lcc}
\hline\hline\noalign{\smallskip}
\multicolumn{1}{l}{Parameter} &
\multicolumn{1}{c}{Model} &
\multicolumn{1}{c}{Model}\\
\multicolumn{1}{l}{} &
\multicolumn{1}{c}{A} &
\multicolumn{1}{c}{B}\\

\noalign{\smallskip}\hline\noalign{\smallskip}
Galactic \nh\ (\ohcm{21}, fixed)      & 0.6                       & 0.6 		      \\  \vspace{1mm}
LMC \nh\  (\ohcm{21})		      & 2.50$_{-0.33}^{+0.19}$    & 2.27$_{-0.22}^{+0.19}$    \\  \vspace{2mm}
n$_{\rm e}$t (10$^{10}$ s cm$^{-3}$)  & 5.7$_{-1.2}^{+1.4}$	  & 5.8$_{-1.3}^{+1.8}$       \\  \vspace{1mm}
He (fixed)			      & 2.57			  & 2.57		      \\  
C (fixed)			      & 0.09			  & 0.09		      \\  \vspace{1mm}
N				      & 0.80$_{-0.27}^{+0.52}$    & 0.95$_{-0.38}^{+0.57}$    \\  \vspace{2mm}
O				      & 0.066$_{-0.013}^{+0.030}$ & 0.099$_{-0.020}^{+0.017}$ \\  \vspace{2mm}
Ne				      & 0.24$_{-0.04}^{+0.07}$    & 0.18$_{-0.02}^{+0.04}$    \\  \vspace{2mm}
Mg				      & 0.27$_{-0.06}^{+0.08}$    & 0.24$_{-0.07}^{+0.04}$    \\  \vspace{2mm}
Si				      & 0.66$_{-0.12}^{+0.16}$    & 0.61$_{-0.12}^{+0.07}$    \\  \vspace{2mm}
S				      & 0.32$_{-0.18}^{+0.20}$    & 0.31$_{-0.16}^{+0.17}$    \\  \vspace{1mm}
Ar (fixed)			      & 0.54			  & 0.54		      \\  
Ca (fixed)			      & 0.34			  & 0.34		      \\  \vspace{1mm}
Fe				      & 0.042$_{-0.009}^{+0.013}$ & 0.058$_{-0.013}^{+0.008}$ \\  \vspace{1mm}
Ni (fixed)			      & 0.62			  & 0.62		      \\
\noalign{\smallskip}\hline
\end{tabular}
\end{center}
\label{pn-fits-b}
\end{table}

For both models there is clear evidence that both the temperature of the soft 
(MEKAL/RAYMOND) component as well as the temperature of the hard (NEI) component have 
increased  after the second epoch (rev.
244) observation. This can also be directly seen in the shape of the EPIC-pn spectra below 1 keV where line intensity 
ratios differ between the spectra from revolutions 244 and 626 (Fig.~\ref{epicpn-spectra}). In 
particular the \ion{Ne}{IX} triplet increased in strength with respect to \ion{O}{VIII}.
The increase in temperature may be related to the progressing equilibration of electron and ion 
temperatures.

The EPIC-pn derived best fit model A reproduces most of the emission lines resolved in the high 
resolution RGS spectra, except the \ion{O}{VII} He-like triplet near 574 eV (Fig.~\ref{rgs-spectra}), 
which suggests that some plasma with temperatures lower than 0.3 keV contributes to the 
emission. Oxygen line ratios derived from the September 2004 Chandra LETG spectrum, taken 
about 1.5 years after the last XMM-Newton observation, are consistent with lower temperatures 
of about 100 eV \citep{2005ApJ...628L.127Z}. 
For a quantitative comparison of the line strengths of the most prominent emission lines 
seen in the LETG and RGS spectra we modeled the RGS spectra in a way similar to the work of 
\citet{2005ApJ...628L.127Z} but with the exception that we modeled both RGS spectra 
simultaneously with an absorbed bremsstrahlung component for the continuum and a set of 
emission lines with Gaussian line profiles. For completeness, we included all those 
emission lines which are detected in the RGS spectra of the bright Magellanic Cloud 
supernova remnants \object{1E\,0102.2$-$7219} \citep{2001A&A...365L.231R} and \object{N132D} 
\citep{2001A&A...365L.242B}. This makes sure that no observed emission line is missed
in the model as is the case for the plasma models used above.
The width of all lines was forced to be the same, i.e. one free parameter. This resulted
in a best fit value for the full width at half maximum of 5.3$\pm$1.0 eV which corresponds to
a Doppler velocity of 2430$\pm$460 \kms.

All line energies were fixed at the laboratory values, but red-shifted by the LMC Doppler shift
of 286 \kms, which was found to be consistent with the LETG spectrum 
\citep{2005ApJ...628L.127Z}. Leaving the common red-shift for all lines as free parameter 
in the fits a Doppler velocity of 460$\pm$126 \kms\ (90\% confidence) is inferred from 
the RGS spectra. Taking into account the RGS wavelength scale calibration uncertainties of  
$\sim$200 \kms\ the LETG and RGS derived LMC Doppler shifts are consistent 
with each other.  The line fluxes derived for the fixed Doppler velocity 
of 286 \kms\ are listed in Table~\ref{lines}. We assumed 286 \kms\ to allow a direct 
comparison with the LETG values published by \citet{2005ApJ...628L.127Z}. Simultaneously we 
verified that the line fluxes change only marginally when the best-fit Doppler shift of 460 \kms\
is used (the fluxes change by less then 5\% which is less than the statistical errors).
As can be seen from the LETG/RGS flux ratios presented in Table~\ref{lines},
all line intensities except the \ion{O}{VII} triplet 
increased over the 1.5 years between the XMM-Newton (May 2003) and Chandra (Sep. 2004)
observations, consistent with the trend seen in the 
EPIC-pn spectra. Although the errors are relatively large there is an indication that the 
ionization degree of O and Ne has increased from the XMM-Newton to the later Chandra observations.

As shown in Table~\ref{chisq} the fit with two ionization equilibrium models, i.e. model D, cannot really be 
excluded based on the values of $\chi^2_r$, but the models A and B involving non-equilibrium models 
appear to fit the data slightly better. In either case the low temperature component is in equilibrium, 
which means that the ionization parameter log(n$\sb e\cdot$t) $>$ 12, whereas it is much lower by about 
a factor of 18 for the high temperature component (log(n$\sb e\cdot$t) $=$ 10.76). 
This could mean that the high temperature component 
has formed predominantly in a region with a  density lower by that factor. 
We note that \cite{2005ApJ...628L.127Z} from the September 2004 LETG observation, which took place 
about 1.5 years later than the XMM-Newton observation discussed here, find the low temperature component in 
equilibrium and the high temperature component out of ionization equilibrium with an ionization parameter 
log(n$\sb e\cdot$t) $=$ 11.23, which is a factor of about three higher than the XMM-Newton value. 
Apart from the separation in observation time the difference between Chandra and 
XMM-Newton here is that the Chandra analysis is restricted to evaluation 
of the emission lines and that \cite{2005ApJ...628L.127Z}
have used a different emission model, which makes use of a continuous distribution of shock velocities with 
the ionization parameter rigidly coupled to the post shock temperature.  
In the following we discuss a possible consequence if the high temperature component would be significantly 
out of ionization equilibrium. 

The region in and close to the equatorial plane of the inner ring is likely 
to have a higher density than the regions well above and below, which suggests 
that the low-temperature component is dominated by emission from the equatorial 
plane, whereas the high-temperature component is preferentially created in regions 
way off the plane. This would be an alternative to the association of the two 
temperatures with the forward and reverse shocks. A very simplified check can be 
made by comparing the emission measures of the two components, which show a ratio 
of 16/1 for model A. If assumed to be disk-like the fractional emission volume 
of the low-temperature component scales with the ratio of the height of the inner 
ring and the radius of the inner ring. We take the ring height to be the same as 
the ring width which can be determined from the early optical/UV image as about 1/15 of 
the radius \citep{1991sos..conf..575P}. Taking the density ratio as above, i.e. 18/1 
and the volume ratio as 1/15 the emission measure ratio is expected to be 
18$\cdot$18/15 = 22/1 in comparison with the measured value of 16/1. Given the 
uncertainties in these numbers, in particular the ring width, and the spatial 
structure, which is certainly present in the matter distribution, the similarity 
between measurement and expectation is encouraging.     

For determining the elemental abundances we followed the procedure applied to the Chandra LETG 
data \citep{2005ApJ...628L.127Z}. Because of the fairly low sensitivity of the X-ray instruments 
values of some abundances were fixed, i.e. for He, C, Ar, Ca, and Ni. The abundances of the remaining 
seven elements N, O, 
Ne, Mg, Si, S and Fe are free parameters for the fits. Our results on the relative abundances   
and the results which  \cite{2005ApJ...628L.127Z} 
derived from the Chandra LETG measurements agree very well, despite the fact that the observations were taken at times 
differing by about 1.5 years. There are, however, two exceptions. We find that the abundance of Si is about 
2 to 3 times higher in the earlier XMM-Newton observations. Either this is  related to the  
models or the Si abundance in the region of the inner ring close to its core is lower than at its inner edge. 
Besides that the XMM-Newton Si abundance (0.54 - 0.82) is also larger than what is generally being quoted 
as the general Si abundance in the LMC (0.31). The second mismatch occurs for Fe, which is found to be 
significantly under-abundant in both measurements, but the LETG derived Fe abundance is at least a factor of about 
two higher than the XMM-Newton result. Does the Fe abundance increase towards the core of the inner ring? 
 
One of the remarkable features of the inner ring is the overabundance of N(1.63) and the under-abundance of 
O(0.18) with a ratio of about 9. This ratio is consistent with the XMM-Newton measurements with the abundances of N 
$\sim$0.8  
and of O $\sim$0.066 yielding a ratio of about 12.1$\pm$6 (c.f. Table~\ref{pn-fits-b}), although the individual values of the N and O  
abundances are both lower by about a factor of 2 to 2.5.    
In general, the abundances of all elements except that of He are lower than solar. But it seems that the seven elements 
which were set free in the fits fall into three groups, which are N and Si in group one with an average 
abundance value of 0.73, Ne, Mg and S in group 2 with an average abundance of 0.28, and O and Fe in 
group 3 with the very low abundance of 0.054. Given the error boundaries there is no overlap between these 
three groups. If this distinction is not a chance coincidence it might tell us something about the creation 
of the inner ring and/or the progenitor. Within each group, though, the
abundance ratio conforms to solar 
values, i.e. the ratio of O/Fe is solar-like, as is that of N/Si, as well as the ratio for the elements in group 
3. It is unlikely that these abundances have resulted from a depletion process, which leaves 
the creation process of the elements in the progenitor star as the most likely origin.

\begin{table}
\caption[]{RGS line fluxes for prominent emission lines.}
\begin{center}
\begin{tabular}{lccc}
\hline\hline\noalign{\smallskip}
\multicolumn{1}{l}{Line} &
\multicolumn{1}{c}{Energy} &
\multicolumn{1}{c}{Flux$^1$} &
\multicolumn{1}{c}{Ratio} \\
\multicolumn{1}{l}{} &
\multicolumn{1}{c}{[eV]} &
\multicolumn{1}{c}{[photons cm$^{-2}$ s$^{-1}$]} &
\multicolumn{1}{l}{LETG/RGS} \\

\noalign{\smallskip}\hline\noalign{\smallskip}
\ion{N}{VII}  &  500         & 29.0$^{+3.1}_{-3.5}$\expo{-6} & 1.53$\pm$0.26 \\
\ion{O}{VII}  &  574/569/561 & 42.9$^{+7.0}_{-7.1}$\expo{-6} & 0.81$\pm$0.19 \\
\ion{O}{VIII} &  654         & 43.5$^{+2.8}_{-3.4}$\expo{-6} & 1.17$\pm$0.13 \\
\ion{O}{VII}  &  666         & ~6.2$^{+2.2}_{-2.2}$\expo{-6} & --            \\
\ion{O}{VIII} &  775         & ~8.9$^{+2.5}_{-1.6}$\expo{-6} & 2.26$\pm$0.69 \\
\ion{Ne}{IX}  &  922/915/905 & 45.4$^{+5.6}_{-5.4}$\expo{-6} & 1.36$\pm$0.19 \\
\ion{Ne}{X}   &  1022        & 19.6$^{+3.1}_{-3.1}$\expo{-6} & 1.86$\pm$0.32 \\
\noalign{\smallskip}\hline
\end{tabular}
\end{center}

$^1$Line fluxes with 1$\sigma$ errors. For the He-like ion triplets the sum of the three lines is
given.
\label{lines}
\end{table}

At last we discuss the column density, the  
LMC \nh\ of (2.50$_{-0.33}^{+0.19}$)\hcm{21} 
for the photo-electric absorption just in the LMC. This can be compared 
with the total column density throughout the LMC, which is measured in \Hone\
to be 2.53\hcm{21} \citep{2005A&A...432...45B}. The agreement means that 
either \sn\ is located at the far side of the LMC or that it shows 
intrinsic absorption by matter in front and close to it. Actually the X-ray measurement
can be used to set an upper limit for cold  
X-ray absorbing matter at distances from the explosion site greater 
than the radius of the optical ring, which is 
$\sim$7\expo{17} cm. If the base of an earlier 
wind of the progenitor star with a r$\sp{-2}$ profile would be located just 
outside the inner ring the upper limit of the base density is 
4270 cm$\sp{-3}$. For a stellar wind with a velocity of 10 km/s this density requires a fairly large 
stellar mass loss rate of 5.6$\times$10$\sp{-4}$ solar masses per year. 
This value is larger by  a factor of about six than the upper end of the mass loss range derived 
from radio observations of type II supernovae 
\citep{2005ASPC..342..290W}. 

In that sense it cannot be excluded that the entire absorption 
is due to circumstellar absorption either from a single star progenitor or a pre-explosion binary 
system. It is curious, though, that this amount of absorption is so close to the LMC total column 
density. But even if the circumstellar absorption would be just a tenth of the total, which would be compatible 
with mass loss rates known for red supergiants, the base of the wind can stand close to the inner ring, 
which would support the suggestion that the ring is a product of the interaction of the winds of the red and 
blue supergiants. Of course structured mass loss in a binary is still not excluded. 
Whether there is such a high density, spherically symmetric shell just outside the inner ring radius 
is likely to be decided by observations of the soft X-ray light-curve over the next couple of years, when the 
shock wave has passed the inner ring.   
  
\begin{figure*}
  \resizebox{12cm}{!}{\includegraphics[angle=-90,clip=]{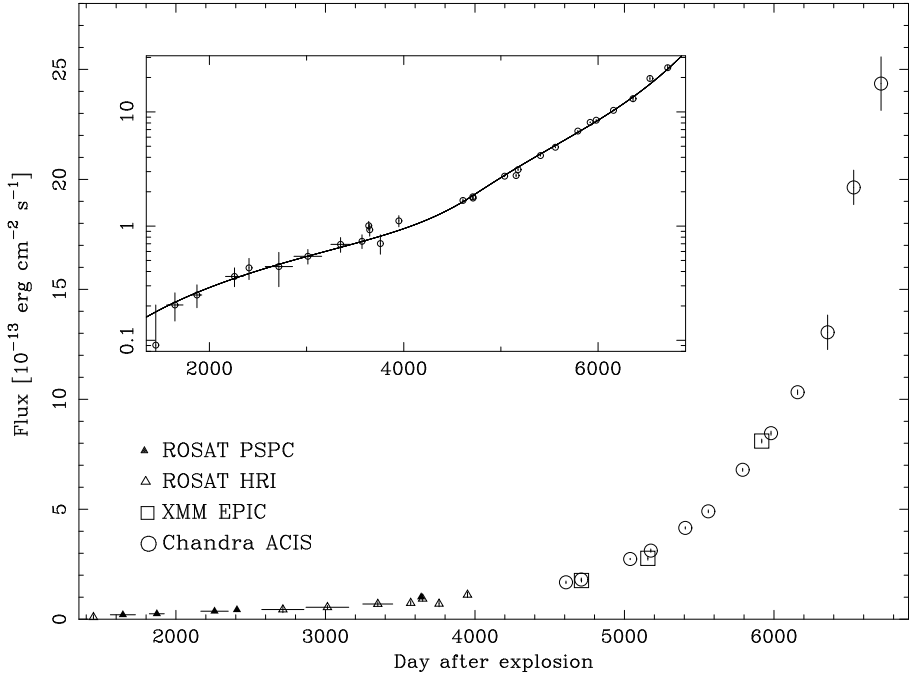}}
  \hfill\parbox[t]{60mm}{
  \vspace{65mm}\caption[]{
      X-ray light curve of \sn\ in the 0.5$-$2.0 keV band. Chandra ACIS-S 
      fluxes were corrected for pile-up losses as described in Sect.~4. The 
      insert shows the light curve in logarithmic intensity scale together 
      with the model described in Sect.~4.
  }
  \label{lcurve}
  }
\end{figure*}

\section{X-ray light curve}

\subsection{Data}

Since the discovery in 1994 the soft X-ray flux has been monitored  first by ROSAT and then  
by Chandra and 
XMM-Newton. \cite{1996A&A...312L...9H} have summarized the early ROSAT data. 
For constructing a self-consistent light curve the energy and flux cross-calibration of the 
various instruments used are needed to be known to a level possibly comparable to the statistical errors. 
In the following we make an attempt to re-calibrate all previous data taken including ROSAT and Chandra. 
Using the best fit model from the EPIC-pn first-light
spectra we re-determined the count rate to flux conversion factor to 
9.05\expo{10} PSPC counts / (erg cm$^{-2}$ s$^{-1}$) and also add the fluxes from five HRI and one PSPC observations performed
up to nearly day 4000 after explosion, which have not been published previously but are in the ROSAT archive. 

Chandra performed a 
regular monitoring of \sn\ since the launch of the observatory with published fluxes before the last 
XMM-Newton observation up to day 5800 
\citep{2004ApJ...610..275P}. Extrapolating the exponential flux increase detected by Chandra
to the epoch of the last XMM-Newton observation we compute a significantly lower flux 
than observed with XMM-Newton. To investigate if this is caused
by a faster brightening of the source or by an instrumental effect we analyzed two more Chandra
observations which are available in the public archive and which were performed after 
the last XMM-Newton observation. We determined count rates in the 0.5$-$2.0 keV band and apply
the same flux conversion factor as \citet{2004ApJ...610..275P} did for their last Chandra 
observation. The two additional flux values follow nicely the extrapolation of the Chandra
light curve demonstrating that the different fluxes given by the two instruments are caused
by a systematic effect. \citet{2004ApJ...610..275P} state in their paper that they did not correct 
the Chandra ACIS fluxes for CCD pile-up effects which they estimate to be $<$10\%.
Pile-up of photons causes flux losses and is therefore the most likely origin for the lower Chandra 
fluxes. In contrast the count rate for the EPIC-pn is far from any pile-up limit, especially 
during the SW mode observation. In addition, pile-up effects should be reduced for the earlier 
observations when the fluxes were lower, consistent with the better agreement between 
XMM-Newton and Chandra fluxes derived from those observations.

We therefore suggest that the lower fluxes derived from Chandra ACIS data are due to pile-up 
and we have corrected the data for the flux losses following the Chandra users handbook 
for the dependence of the pile-up fraction on the photon rate with a normalization using 
the last XMM-Newton observation as reference. We find a pile-up correction for day 6158 
of about 22\%, which can clearly not be neglected. Three more Chandra fluxes were published 
by \citet{2005ApJ...634L..73P}. During those observations the CCD readout time was reduced 
by a factor of about two to reduce pile-up effects. However, for the last Chandra 
observations the intensity increased by more than a factor of two compared to day 6158
and therefore results in similar pile-up losses. We corrected also these last three Chandra fluxes
as before, taking also into account the reduced count rate per readout frame.

\begin{table}
\caption[]{X-ray fluxes of \sn.}
\begin{center}
\begin{tabular}{llcc}
\hline\hline\noalign{\smallskip}
\multicolumn{1}{l}{Age} &
\multicolumn{1}{l}{Instrument} &
\multicolumn{1}{c}{0.5-2.0 keV Flux} &
\multicolumn{1}{c}{Pile-up} \\
\multicolumn{1}{l}{[days]} &
\multicolumn{1}{l}{} &
\multicolumn{1}{c}{[\oergcm{-13}]} &
\multicolumn{1}{c}{correction} \\

\noalign{\smallskip}\hline\noalign{\smallskip}
1448  & ROSAT HRI    &  0.091$\pm$0.113 & - \\
1645  & ROSAT PSPC   &  0.204$\pm$0.057 & - \\
1872  & ROSAT PSPC   &  0.250$\pm$0.057 & - \\
2258  & ROSAT PSPC   &  0.363$\pm$0.068 & - \\
2408  & ROSAT PSPC   &  0.431$\pm$0.091 & - \\
2715  & ROSAT HRI    &  0.442$\pm$0.147 & - \\
3013  & ROSAT HRI    &  0.544$\pm$0.079 & - \\
3350  & ROSAT HRI    &  0.69$\pm$0.10   & - \\
3570  & ROSAT HRI    &  0.74$\pm$0.10   & - \\
3640  & ROSAT PSPC   &  1.01$\pm$0.09   & - \\
3650  & ROSAT HRI    &  0.93$\pm$0.11   & - \\
3760  & ROSAT HRI    &  0.70$\pm$0.14   & - \\
3950  & ROSAT HRI    &  1.11$\pm$0.12   & - \\
4609  & Chandra ACIS &  1.68$\pm$0.06   & 1.035 \\
4711  & Chandra ACIS &  1.81$\pm$0.07   & 1.038 \\
4712  & XMM EPIC-pn  &  1.76$\pm$0.07   & - \\
5038  & Chandra ACIS &  2.74$\pm$0.03   & 1.057 \\
5156  & XMM EPIC-pn  &  2.77$\pm$0.08   & - \\
5176  & Chandra ACIS &  3.12$\pm$0.07   & 1.065 \\
5407  & Chandra ACIS &  4.15$\pm$0.05   & 1.087 \\
5561  & Chandra ACIS &  4.91$\pm$0.07   & 1.102 \\
5791  & Chandra ACIS &  6.79$\pm$0.07   & 1.142 \\
5918  & XMM EPIC-pn  &  8.10$\pm$0.09   & - \\
5980  & Chandra ACIS &  8.46$\pm$0.09   & 1.177 \\
6158  & Chandra ACIS & 10.32$\pm$0.11   & 1.216 \\
6359  & Chandra ACIS & 13.05$\pm$0.78   & 1.136 \\
6533  & Chandra ACIS & 19.64$\pm$0.79   & 1.205 \\
6716  & Chandra ACIS & 24.35$\pm$1.22   & 1.255 \\
\noalign{\smallskip}\hline
\end{tabular}
\end{center}
\label{tab-lcurve}
\end{table}

The final, corrected X-ray light curve in the 0.5$-$2.0 keV band is presented in Fig.~\ref{lcurve} 
with the fluxes listed in Table~\ref{tab-lcurve}. The flux changes linearly with time until about 
day 3500 and continues to increase exponentially until day 6700. The insert shows the light curve 
on a lin-log scale together with the prediction constructed from a model of the matter distribution 
inside the volume run over by the blast wave. 

So far we have concentrated on the soft X-ray supernova light curve but Chandra and XMM-Newton also allow to derive 
a hard X-ray light curve for the band 3 - 10 keV. The three XMM-Newton data points on days 4712, 5156 and 5918 agree very well 
with the Chandra data points given by \cite{2005ApJ...634L..73P} and confirm the fairly slow increase of the hard flux compared to the 
soft emission increase up to day  5918.  

\subsection{Model}

The emission from \sn\ is obviously thermal, so that the observed flux is proportional to the 
cooling function and the emission measure and inversely proportional to the distance squared. 
For the distance we take 50 kpc. The value of the cooling function is taken from the observational best 
fits to the spectra; the observed temporal changes in temperature have a very 
small effect on the cooling function in the energy band of 0.5-2 keV, so that we take 
the cooling function in the band to be constant in time for the observations at hand. 
Therefore the shape of the \sn\ soft X-ray light curve is determined 
by the emission measure, i.e. the matter density squared integrated over the 
volume that has been encompassed  by the forward shock or blast wave. 

\begin{figure}
\resizebox{0.95\hsize}{!}{\includegraphics[angle=-90,clip=]{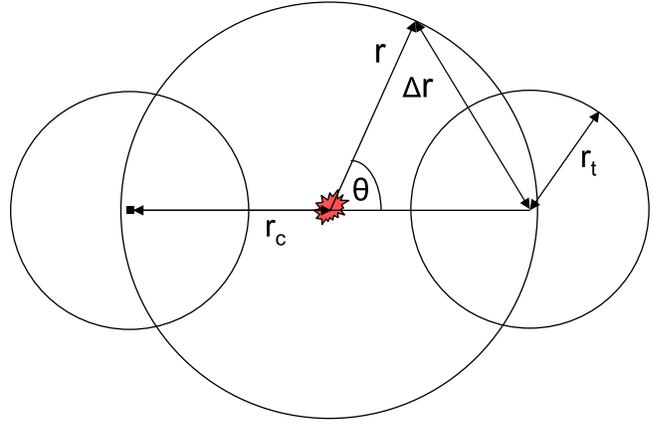}}
\caption{Sketch used to model the matter distribution around \sn. The matter is arranged in a circular 
torus of radius r$\sb c$; lines of constant density form circles which build up the torus. 
For torus circles up to radius r$\sb t$ the distribution follows a Gaussian which has its 
peak at r$\sb c$ and falls off with increasing $\Delta$r. Outside of r$\sb t$ the distribution is 
roughly exponential which peaks at r$\sb t$ and decreases with increasing 
$\Delta$r. For each location in space (r, $\Theta$, $\phi$) the density is defined according to 
this scheme (no dependence on $\phi$). The big circle represents the radius of the forward shock wave at 
day 7000, which has just passed the torus center.}
\label{model_sketch}
\end{figure}

The key element is to find a reasonable description for the spatial distribution of the matter 
density. We assume the inner ring to be a circular torus  with a radius of r$\sb c$ measured to be 
{6.2$\times$10$\sp{17}$ cm}
(0.83\arcsec) \citep{1991sos..conf..575P}, a circular cross section  and the  origin 
coincident with the explosion site.  
The density distribution is independent of the azimuth along the torus perimeter - it is circular 
symmetric. Within the torus the density n$\sb{\rm H}$ is described by a Gaussian, i.e. 
n$\sb{\rm H}~\propto$~exp[-($\Delta\rm r/\rm h\sb i$)$\sp 2$], with $\Delta\rm r$ being 
the shortest distance from the torus center line, not the torus origin, though. At some torus radius r$\sb t$ a 
transition from the Gaussian form to a simple exponential dependence occurs, 
i.e. 
n$\sb{\rm H}~\propto$~exp[-($\Delta\rm r/\rm h\sb o$)] for $\Delta\rm r \ge$~r $\sb t$ 
(see Fig.~\ref{model_sketch}). 
The exponential dependence is further modified by a factor of 1/r where r is the distance from the 
origin. The two forms are normalized to each other such that a smooth transition in density occurs 
at $\Delta\rm r=$~r$\sb t$. In addition to h$\sb i$, h$\sb o$ and r$\sb t$, which are to be 
determined, the location of the blast wave front r$\sb e$ at some specified epoch is a free parameter. 
For this paper we assume the blast wave velocity to be constant in time up to day 7000. 
The shocked matter or the X-ray emitting volume elements, are assumed to expand radially and adiabatic  
with a speed of three quarters of the blast wave velocity. 

An acceptable fit to the data (c.f. Fig.~\ref{lcurve}) has been obtained for r$\sb e$ = 
6.3$\times$10$\sp{17}$ cm (0.84\arcsec) at day 7000, 
which means that the forward shock has just passed the center of the torus.
This also means that the mean blast wave shock velocity over the past 7000 days is 10400 \kms\
for a distance of 50 kpc.   
The transition radius  r$\sb t$ = 
3.6$\times$10$\sp{17}$ cm (0.48\arcsec), 
so that the region inside the 
torus in which the Gaussian density distribution prevails 
is roughly half of the torus radius r$\sb c$. The scale height 
is h$\sb i$ = 2.1$\times$10$\sp{17}$ cm (0.28\arcsec). The scale height of the exponential ``atmosphere" 
of the ring is h$\sb o$ = 
2.6$\times$10$\sp{17}$ cm
 (0.35\arcsec).  
From adjusting the measured and the best-fit fluxes we get a density of 
n$\sb{\rm{H,c}}$~=~1.15$\times$10$\sp 4$~cm$\sp{-3}$ for the innermost core of the ring. This number can be compared 
with the ring electron density of n$\sb{\rm e}$~=~(2~-~4)$\times$10$\sp 4$~cm$\sp{-3}$
 derived from the early optical/UV measurements by \cite{1989ApJ...336..429F}. The ring width, which shows 
up in these optical/UV measurements, is 
6.7$\times$10$\sp{16}$ cm
 ($~$0.09\arcsec) \citep{1991sos..conf..575P} and \cite{1989ApJ...336..429F} 
estimate for the included mass 0.03~-~0.05~M$_{\odot}$. Our model gives for that limited region a 
mass of 0.065 solar masses, which is slightly higher. But this number is very sensitive on the exact width of 
the ring because of the Gaussian density dependence. The total mass overrun by the forward shock wave 
is, however, much greater and amounts to 0.45~M$_{\odot}$, with 90\% \ of the mass contained 
in the much wider ring and 10\% \ in the exponential ``atmosphere". The total mass around \sn\ out to day 7000 
already exceeds the previously quoted mass of the inner ring by a factor of 10, which may have some relevance to the models 
explaining the origin of the ring, as far as the energetics are concerned. If the upstream density 
distribution beyond day 7000 follows the functional forms given above, a total mass of 1.3~M$_{\odot}$ will 
be involved. The soft X-ray light curve is predicted to flatten in mid 2006 and will reach its maximum 
luminosity of 2$\times$10$\sp{36}$ erg/s in early 2009. 

Unacceptable fits to the data, either in the early part or in the late part of the light curve, 
become apparent for changes of r$\sb e$, r$\sb t$ and h$\sb i$ exceeding 5 to 7\%. This error range is 
so small basically because r$\sb c$ is fixed and the distribution is steep. 
This is different for h$\sb o$, the scale height of the 
early data; h$\sb o$ can be changed up to factor of 2 until deviations become significant. The late 
time light curve is not affected. Within this range of errors the central density changes by up to 
30\%.

The numbers given above have been derived for a constant blast wave velocity v$\sb b$, which is 
certainly not the case according to the latest Chandra results \citep{2004ApJ...610..275P}. This will affect 
our estimates of n$\sb{\rm H,c}$ and h$\sb i$, as well as h$\sb o$, 
 because they scale linearly proportional to v$\sb b$ and 
inversely proportional to v$\sb b$, respectively. 
The slow down of  v$\sb b$ reaches values of up to  
a factor of 2 to 3, in the late stage of the expansion, and accordingly  
h$\sb i$ is expected be lower and n$\sb{\rm{H,c}}$ might reach (2 - 4)$\times$10$\sp 4 $cm$\sp{-3}$ on the average. 
In contrast, the estimates of the mass in the ring is only slightly affected and will stay close to the 
large values given. In a separate paper we will describe the model in more detail and will present 
results for a model with v$\sb b$ slowing down in time.

\section{Limits on a compact object}

\subsection{Observations}

Up to now the search for a compact remnant has not revealed any 
significant evidence for the existence of a neutron star or black hole, which may be due to 
various reasons. If a pulsar with a wind nebula were created in the explosion one would expect to see some 
power-law like X-ray spectrum photo-electrically absorbed by the ambient debris of the 
progenitor's envelope. Depending on the column density the pulsar and its nebula should appear at some time. 
To overcome the absorption one should look for the emission in the hard X-ray regime first, and XMM-Newton 
can supply some data and clues. 

To estimate the putative contribution from a compact object to the total flux a power-law component
with fixed photon index (2.1) was added to model A in the combined fit to the EPIC-pn spectra. 
The intensity was not allowed to vary with time, i.e. it was constrained mainly by the first-light 
spectra in January 2000. The $\chi^2$ improved only to 803.5 which formally does not justify 
the additional free parameters. For 90\% confidence an upper limit for the 
flux in the power-law component can be derived. For the 0.2$-$10.0 keV (0.5$-$2.0 keV) band 
this results in an observed, i.e. not corrected for absorption, flux of 1.1\ergcm{-13} (4.4\ergcm{-14}) 
which corresponds to an absorption corrected luminosity 
 of 5.7\ergs{34} (2.1\ergs{34}) for a distance of 50 kpc.
This is slightly less than the absorption uncorrected 2.3\ergs{34} upper limit in 
the Chandra images 
\citep[0.5$-$2.0 keV, ][]{2000ApJ...543L.149B}.
Even if a pulsar was created in the explosion and even if the stellar debris is fairly 
transparent in the meantime there 
are other reasons why the pulsar has not emerged in the observations. 

\subsection{Discussion}
\label{lab-discussion}

The detection of neutrinos immediately after the appearance of \sn\
is taken as proof that in this supernova a (proto-) neutron star was  
created.  Assuming magnetic flux and angular momentum conservation 
during the collapse a dipolar surface field of the order of $10^{12}$ G 
and a rotational  period not much smaller than 0.01~s are expected.
Assuming the ``standard'' parameter for the moment of 
inertia $I = 10^{45}$ g~cm$^2$, we calculate the  spin-down luminosity 
$L_{sd} = 4\pi^2 I \dot{P}/P^3$ $\approx$ $4\times10^{52}\,\dot{P}$ erg s$^{-1}$. 
 Since the X-ray luminosity $L_x = 10^{-3}\,L_{sd}$ 
\citep{1997A&A...326..682B, 2005ApJ...subm....G},
we find a lower limit of $\dot{P} \ge 10^{-15}$ in order   
for the expected X-ray luminosity of the pulsar to significantly exceed 
the observed upper limit. If the   
rotational energy loss of the pulsar is due to magneto-dipole radiation the  
dipolar surface magnetic field $B_d = 2\times 10^{12}\,(P \dot{P}_{-15})^{1/2}$ G,
where again the ``standard'' parameters have been applied, and $ 
\dot{P}_{-15} = \dot{P}/10^{-15}$ s s$^{-1}$. 

In order to get the X-ray 
luminosity of the pulsar larger than $5.7\times 10^{34}$ erg s$^{-1}$, 
i.e. to enforce a spin-down faster than $\dot{P}_{-15} = 1$,   
the surface dipolar field strength $B_d > 2 \times 10^{12}\,\sqrt{P}$ G, 
i.e. larger than $2 \times 10^{11}$ G. If the field of the pulsar is lower it would still 
escape current detection capabilities. Furthermore,  in case of \sn,  
 matter fall-back might have been so powerful that approximately 0.1~M$_{\odot}$
have been accreted 
\citep{1989ApJ...346..847C}, 
which would submerge the initially existing pulsar magnetic fields down to  
regions close to the crust-core interface of the neutron star.  
Depending on the equation of state and on the conductive properties  
of the crust it takes at least 1000 years if not much longer, until  
the field is re-diffused to the surface with a strength  
larger than $10^{11}$ G 
\citep[see][ and references therein]{1999A&A...345..847G}.

The only other possibility we see for the neutron star in \sn\ to have a  
surface magnetic field exceeding $2\times 10^{11}$ G is the  
transformation of thermal into magnetic energy via the strong  
temperature gradient present in the liquid crust of the young star  
causing a thermo-magnetic instability 
\citep{1986MNRAS.219..703U,1996A&A...309..203W}.
The typical period to reach poloidal surface  
fields of the order of $10^{12}$ G however, exceeds certainly $\sim  
100$ yrs. In conclusion,  a pulsar powered signature in the spectrum of \sn\
is not very likely to appear very soon.

\section{Conclusions}

We present the results from a spectral analysis of XMM-Newton data from
\sn\ obtained in January 2000, April 2001 and May 2003. The spectra are
consistent with thermal emission from the supernova remnant without a
significant hard power-law component (with an absorption corrected 
0.2-10.0 keV luminosity limit of 5.7\ergs{34}) originating from a 
compact object.
Modeling the spectra with two components involving emission from plasma
in collisional ionization equilibrium and/or in non-equilibrium yields
clear evidence for a temperature rise after the second epoch observation
for both components, reaching $\sim$0.3 keV and $\sim$3 keV in the soft 
and hard component, respectively. 

We have collected all available X-ray 
fluxes in the 0.5-2.0~keV band from ROSAT, XMM-Newton and Chandra for the 
light curve of \sn\ up to day 6716 after the explosion. This includes 
previously unpublished ROSAT data available in the archive and published
Chandra fluxes which we corrected for photon pile-up effects present in 
the CCD spectra. The final light curve shows a linear increase up to 
about day 4000 and an exponential rise afterwards. We model the light 
curve with emission from a circular torus around the explosion site which
represents the inner ring around \sn. From the best fit we find that the 
forward shock just passed the center of the torus and that the mean 
shock velocity over 7000 days is 10400 \kms. With a core density of the 
torus of 1.15$\times$10$^4$ cm$^{-3}$ we find a total mass overrun by the 
forward shock of 0.45~M$_{\odot}$. The model, assuming a constant blast 
wave velocity in its simplest form, predicts a flattening in the X-ray 
light curve for mid 2006. A new XMM-Newton observation is 
currently scheduled for the end of 2006 to further monitor the spectral
evolution of \sn.

\begin{acknowledgements}
The XMM-Newton project is an ESA Science Mission with instruments
and contributions directly funded by ESA Member States and the
USA (NASA). The XMM-Newton project is supported by the
Bundesministerium f\"ur Wirtschaft und Technologie/Deutsches Zentrum
f\"ur Luft- und Raumfahrt (BMWI/DLR, FKZ 50 OX 0001), the Max-Planck
Society and the Heidenhain-Stiftung. 
\end{acknowledgements}

\bibliographystyle{aa}
\bibliography{general,myrefereed,ins,ism}

\end{document}